\documentclass[conference]{IEEEtran}
\usepackage{cite}
\usepackage{amsmath,amssymb,amsfonts}
\usepackage{algorithmic}
\usepackage{graphicx}
\usepackage{textcomp}
\usepackage{xcolor}
\usepackage{url}
\usepackage{booktabs}
\usepackage{comment}
\makeatletter
\newcommand{\linebreakand}{%
  \end{@IEEEauthorhalign}
  \hfill\mbox{}\par
  \mbox{}\hfill\begin{@IEEEauthorhalign}
}
\makeatother

\def\BibTeX{{\rm B\kern-.05em{\sc i\kern-.025em b}\kern-.08em
    T\kern-.1667em\lower.7ex\hbox{E}\kern-.125emX}}
\begin{document}

\title{Mind Companion: An Embodied Conversational Agent for Process-Based Psychotherapy
}

\author{
\IEEEauthorblockN{Sofie Kamber}
\IEEEauthorblockA{\textit{ETH Zurich}\\Switzerland\\skamber@ethz.ch}
\and
\IEEEauthorblockN{Lukas Diebold}
\IEEEauthorblockA{\textit{ETH Zurich}\\Switzerland\\ldiebold@ethz.ch}
\and
\IEEEauthorblockN{Pascal Riachi}
\IEEEauthorblockA{\textit{ETH Zurich}\\Switzerland\\priachi@ethz.ch}
\linebreakand
\IEEEauthorblockN{\strut Stella Brogna}
\IEEEauthorblockA{\textit{University of Lucerne}\\Switzerland\\stella.brogna@unilu.ch}
\and
\IEEEauthorblockN{\strut Andrew Gloster}
\IEEEauthorblockA{\textit{University of Lucerne}\\Switzerland\\andrew.gloster@unilu.ch}
\and
\IEEEauthorblockN{\strut Rafael Wampfler}
\IEEEauthorblockA{\textit{ETH Zurich}\\Switzerland\\rafael.wampfler@inf.ethz.ch}
}

\maketitle

\begin{abstract}
Access to evidence-based psychotherapy remains limited worldwide, with long waitlists even in high-income regions. Recent advances in large language models (LLMs) offer potential for scalable mental health support when designed with clinical oversight and safety mechanisms. We present Mind Companion, an LLM-based embodied conversational agent integrating multi-layered psychological analysis with process-based therapy principles. The system performs real-time analysis of client statements across fact extraction, psychological flexibility process detection, emotion recognition, and safety monitoring. Analysis results are stored for supervising clinicians to inform therapeutic planning. Response generation incorporates retrieval-augmented generation from evidence-based therapeutic literature and context-aware prompting. Responses are delivered through an embodied avatar with synchronized speech synthesis and animation. We evaluated three LLM configurations (GPT-4.1-mini, GPT-5.2, Claude Sonnet 4.5) against therapist responses from real therapy sessions using automated LLM-judge assessment and expert evaluation with 11 professional psychotherapists. GPT-5.2 achieved higher ratings than human therapist responses across understanding, interpersonal effectiveness, collaboration, and therapeutic alignment in both evaluations, demonstrating the feasibility of LLM-based conversational agents as tools to complement clinical care.
\end{abstract}


\section{Introduction}

Mental health disorders are among the leading causes of disability worldwide.
The Global Burden of Disease 2019 study estimated 125.3 million disability-adjusted life years (DALY) --- approximately 4.9\% of all DALYs --- attributable to mental disorders in 2019, up from 80.8 million in 1990~\cite{GBD2019MentalDisorders2022}.
Despite this need, access to evidence-based psychotherapy remains severely constrained: the WHO Mental Health Atlas 2020 reports a global median of only 13 mental health workers per 100{,}000 people, with long waitlists even in high-income regions such as Europe, where waiting times often exceed two months~\cite{WHO2021MentalHealthAtlas,EU2016WaitingPsychotherapy}.
Scalable digital tools that complement clinical care offer a promising avenue to bridge this gap.

Recent advances in large language models (LLMs) point to a possible path forward. LLMs now achieve near-expert performance on benchmark medical tasks~\cite{Singhal2025MedPaLM2}, and chatbot responses have been rated higher than physician responses for quality and empathy~\cite{Ayers2023JAMA}.
LLM-based chatbots incorporating evidence-based therapeutic frameworks such as Acceptance and Commitment Therapy (ACT) have demonstrated clinically meaningful improvements in well-being among both adults and adolescents~\cite{Naor2022Kai,Vertsberger2022KaiAdol}.

However, significant concerns remain regarding autonomous conversational agents in mental health contexts. Recent research has documented critical failure modes in LLM responses, including inadequate handling of crisis situations such as suicidal ideation and inappropriate responses to severe mental health states including delusions and mania~\cite{moore_expressing_2025}.
Furthermore, existing systems typically lack multimodal expressive capabilities (voice, facial animation, gesture) that could enhance therapeutic presence and alliance-building. These limitations highlight the need for clinically grounded systems combining expressive agent behavior with explicit safety monitoring and clinician-in-the-loop oversight.

We present Mind Companion, an LLM-based embodied conversational agent designed to address these gaps. The system integrates multi-layered real-time analysis with process-based therapy principles to provide a tool for clinical settings rather than autonomous intervention.
Our approach combines four parallel analysis streams --- fact extraction, psychological flexibility process detection, emotion recognition, and safety monitoring --- with response generation guided by retrieval-augmented generation from evidence-based therapeutic literature.
Analysis results are stored and accessible to supervising clinicians, providing structured summaries that inform therapeutic planning and decision-making.
Responses are delivered through an embodied avatar with synchronized speech synthesis and facial animation.

We evaluate the system through two complementary approaches. First, automated assessment compares three LLM backends (GPT-4.1-mini, GPT-5.2, Claude Sonnet 4.5) against human therapist responses from real ACT therapy sessions using LLM-as-judge methodology.
GPT-5.2 performed best and significantly outperformed human therapist responses across understanding, collaboration, interpersonal effectiveness, ineffectiveness, and ACT alignment.
Second, expert evaluation with 11 professional psychotherapists validated that GPT-5.2 outperformed human therapist responses across all dimensions.
Together, these results demonstrate a practical approach to leveraging LLM capabilities for mental health support while addressing critical concerns around safety, clinical grounding, and therapeutic presence essential for responsible deployment in healthcare settings.

\subsection{Contributions}

Our work makes the following contributions:

\begin{itemize}
\item An embodied conversational agent architecture integrating multi-layered real-time psychological analysis (facts, psychological flexibility processes, emotions, safety monitoring) with process-based therapy principles for supervised clinical deployment.
\item A response generation approach combining retrieval-augmented generation from evidence-based therapeutic literature with stage-dependent prompting adapting between assessment and intervention.
\item Evaluation with professional psychotherapists demonstrating that LLM-generated responses can match or exceed human therapist responses on verbal content dimensions in isolated response assessment, while emphasizing the continued need for clinical oversight.
\item A mobile application with embodied avatar presentation featuring synchronized speech synthesis and animation.
\end{itemize}

\section{Related Work}

\subsection{Process-Based Therapy Frameworks}

Recent advances in clinical psychology have shifted from disorder-specific protocols toward process-based therapy (PBT), which focuses on identifying and targeting core psychological processes~\cite{Hayes2019PBT}. PBT's emphasis on flexible intervention strategies makes it particularly amenable to computational implementations, as psychological processes can be more readily detected in conversational data than broad diagnostic categories.

Within this framework, ACT has emerged as a prominent empirically supported instantiation.
ACT is a contextual cognitive behavioral therapy aimed at increasing psychological flexibility through six interconnected processes: acceptance, cognitive defusion, present-moment awareness, self-as-context, values clarification, and committed action~\cite{Hayes2006ACT, hayes2011acceptance}.
Research demonstrates that ACT achieves outcomes comparable to Cognitive Behavioral Therapy (CBT) across mental and physical health conditions~\cite{gloster2020empirical}.

Our work adopts ACT as the therapeutic framework, operationalizing its six processes for automated detection and integrating ACT principles into response generation through retrieval from evidence-based literature.

\subsection{Large Language Models in Digital Mental Health}

LLMs have been increasingly deployed as therapeutic conversational agents with demonstrated clinical effectiveness. Woebot significantly reduced depression symptoms in college students~\cite{fitzpatrick2017delivering}, while Wysa demonstrated well-being improvements and therapeutic alliance development~\cite{beatty2022evaluating}. The iCare system implemented a hybrid architecture combining RASA with LLMs as a fallback mechanism~\cite{icare}.
ACT-based chatbots like Kai.ai reported clinically meaningful improvements in WHO-5 scores among adults and adolescents~\cite{Naor2022Kai,Vertsberger2022KaiAdol}. In addition, retrieval-augmented generation (RAG) has emerged as a key technique in healthcare AI~\cite{xiong2024benchmarking,xiong2024improving}, reducing hallucinations and improving accuracy by grounding LLM outputs in authoritative literature~\cite{yang2024retrieval}.
However, significant concerns remain regarding autonomous mental health chatbots. Recent research documented multiple failure modes, including inadequate handling of crisis situations and inappropriate responses to severe mental health states~\cite{moore_expressing_2025}, underscoring the need for explicit clinical oversight.
LLMs have also been utilized to analyze therapeutic processes systematically. Recent work introduced an LLM-based pipeline for detecting psychological processes and generating process networks from therapy transcripts with high clinical utility~\cite{ong_using_2025}. 

Our work integrates real-time process detection with RAG from evidence-based ACT literature, combines multiple analysis streams (facts, processes, emotions, safety) to inform response generation and clinician oversight, and addresses safety concerns through explicit crisis detection.

\subsection{Embodied Agents in Therapeutic Contexts}

Embodied conversational agents (ECAs) add social presence through gaze, facial expressions, and vocal prosody, influencing user engagement, disclosure, and alliance-building~\cite{provoost2017embodied,tremain2020therapeutic}.
Research shows embodied presentation enhances therapeutic interactions~\cite{provoost2017embodied} and may be particularly effective for addressing momentary emotional states~\cite{balan2024use}.

Users can develop therapeutic alliance with virtual agents comparable to human therapists. The SimSensei virtual interviewer combined multimodal sensing with conversational skills to detect distress and encourage communication~\cite{Gratch2014Ellie,DeVault2014DialogDemo}. Alliance quality predicts engagement and outcomes~\cite{tremain2020therapeutic,beatty2022evaluating}. ECAs embedded in preventive interventions like unguided iCBT have shown feasibility and acceptability~\cite{Suganuma2018ECA}.

Recent advances in audio-driven avatar generation enable streaming speech-to-avatar synthesis with low latencies~\cite{prabhune2023towards}. Systems like NVIDIA Audio2Face generate synchronized lip movements and emotional expressions from speech prosody.

Our work combines embodied presentation with LLM-based response generation, enabling conversational flexibility and therapeutic adaptability while maintaining multimodal engagement benefits.

\section{Method}

\subsection{System Requirements}

The design of \textit{Mind Companion} addresses key challenges of deploying LLM-based conversational agents in therapeutic contexts with clinical oversight:

\begin{enumerate}
\item \textit{Real-Time Responsiveness:} Maintain low latency to enable natural turn-taking through parallel analysis execution and synchronized speech and facial animation.

\item \textit{Multi-Layered Analysis:} Extract facts, psychological flexibility processes, emotions, and safety indicators in real-time to inform response generation and provide structured summaries for supervising clinicians.

\item \textit{Safety Monitoring:} Detect crisis situations (suicidal ideation, self-harm) and conditions where LLM interaction may be counterproductive (delusions, mania) with explicit intervention protocols.

\item \textit{Clinical Grounding:} Ground responses in evidence-based ACT literature through retrieval-augmented generation to minimize hallucinations and maintain therapeutic alignment.

\item \textit{Multimodal Presentation:} Deliver synchronized voice and expressive 3D avatars to enhance therapeutic presence and engagement.

\item \textit{Modular Architecture:} Supports future integration with clinical workflows.
\end{enumerate}

\begin{figure*}
  \centering
  \includegraphics[width=\textwidth, trim=0.8cm 8.3cm 2cm 2.5cm, clip]{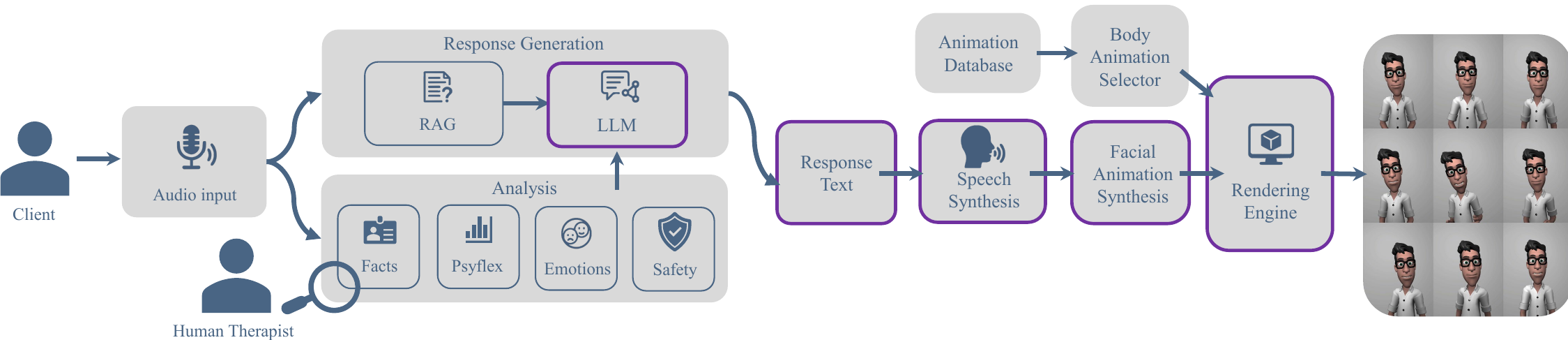}
  \caption{System Overview: Client speech is first transcribed to text, then processed through multi-layered analysis (facts, psychological flexibility processes, emotions, safety) that executes in parallel with response generation for real-time interaction. Analysis results inform LLM-based response generation with retrieval-augmented generation (RAG) from evidence-based ACT literature and provide structured information for therapist oversight. Generated responses are delivered through embodied presentation combining speech synthesis with synchronized 3D avatar animation (body and facial) via the Unity client. Components with a purple border indicate streaming modules that enable low-latency, natural turn-taking.}
  \label{fig:system_architecture}
\end{figure*}

\subsection{System Overview}

We developed an embodied conversational agent that integrates multi-layered psychological analysis with ACT-aligned response generation (see Figure~\ref{fig:system_architecture}). The system is designed for deployment in therapeutic settings with clinician oversight rather than as a standalone autonomous intervention, addressing documented safety and ethical concerns with fully autonomous mental health AI systems~\cite{moore_expressing_2025, iftikhar_how_2025}. The modular design enables future integration with clinical workflows (addressing requirement 6).

The architecture comprises three integrated components:

\begin{itemize}
    \item User Analysis: Multi-layered processing of each client statement to extract facts, detect psychological flexibility processes, recognize emotions from speech, and monitor safety (addressing requirements 2 and 3). Analysis components execute in parallel with response generation to maintain real-time responsiveness (addressing requirement 1). Analysis results are stored for therapist access and used to inform response generation.
    
    \item Response Generation: Context-aware prompting that incorporates analysis results and retrieval-augmented generation (RAG) from ACT literature to produce therapeutically appropriate responses (addressing requirement 4).
    
    \item Embodied Presentation: Multimodal output rendering combining text-to-speech synthesis with synchronized avatar animation to create natural, engaging interactions (addressing requirement 5).
\end{itemize}

\subsection{User Analysis}
\label{subsec:user_analysis}

To maintain conversational naturalness while meeting requirement 1 (real-time responsiveness), all analysis components execute in parallel with response generation. When a user completes an utterance, the system immediately begins generating a response while simultaneously processing the statement through multiple analysis pipelines. This architecture ensures real-time interaction at the cost of a one-turn delay: analysis results from turn $t$ inform response generation at turn $t+1$. 

All user analysis components use GPT-4.1-mini via Azure OpenAI Service for efficiency and cost-effectiveness, unless otherwise stated. Extracted facts and analysis results are stored in a SQLite database for efficient retrieval and persistence across sessions. Table~\ref{tab:analysis-prompts} summarizes the structure of prompts used across the four analysis components. Each prompt follows a consistent architecture: establishing the task and role, defining key concepts or categories, providing guidelines for classification or extraction, and specifying the output format. This structured approach ensures consistent, interpretable outputs that can inform both response generation and clinician oversight.

\begin{table*}
\centering
\caption{Prompt structure for user analysis components. All components use structured prompts with consistent elements to ensure reliable, interpretable outputs.}
\label{tab:analysis-prompts}
\resizebox{\textwidth}{!}{%
\begin{tabular}{p{0.16\linewidth}p{0.19\linewidth}p{0.19\linewidth}p{0.19\linewidth}p{0.19\linewidth}}
\toprule
\textbf{Component} & \textbf{Fact Extraction} & \textbf{Process Detection} & \textbf{Emotion Recognition} & \textbf{Safety Monitoring} \\
\midrule
\textbf{Task Definition} & Extract therapeutically relevant objective facts & Classify ACT psychological processes & Analyze emotional state from text & Identify crisis situations and psychosis \\
\addlinespace
\textbf{Input Modality} & Current user utterance (text) & Current user utterance (text) & Current user utterance (text) & Current user utterance (text) \\
\addlinespace
\textbf{Key Concepts} & 8 fact categories (objective only) & 12 ACT processes (6 flexible, 6 inflexible) & 6 basic emotions (Ekman) & 2 risk categories (crisis, unsafe-to-interact) \\
\addlinespace
\textbf{Guidelines} & Exclude subjective states; capture patterns only; max 100 facts total & Conservative labeling; require clear evidence; max 100 processes total & Infer emotion from semantic content & Distinguish active symptoms from past/managed conditions \\
\addlinespace
\textbf{Output Format} & JSON: new facts, updates, categories & JSON: process labels, explanations, intensity & JSON: 6 emotion scores (0-1) & JSON: risk score, category, reason \\
\bottomrule
\end{tabular}
}
\end{table*}

\subsubsection{Speech-to-text processing}

Client speech is converted to text using Azure Speech-to-Text, providing the foundation for all downstream analysis.

\subsubsection{Fact extraction}

The fact extraction component identifies and maintains a structured knowledge base of relevant client information. For each new user statement, the system analyzes the current user utterance along with the existing fact database to determine whether new facts should be added, existing facts updated, or outdated information removed. To maintain performance and relevance, the system retains a maximum of 100 facts in total across all categories.

Facts are organized into eight clinically relevant categories: \textit{name}, \textit{gender}, \textit{age}, \textit{occupation}, \textit{living context}, \textit{relationships}, \textit{life events}, and \textit{medical history}. The system is designed to capture only objective, therapeutically relevant information --- patterns and significant circumstances rather than one-time routine activities. For example, ``Sleeps 3--4 hours per night'' captures a health-relevant pattern, while ``Watched TV yesterday'' would be excluded as a trivial daily event.

All extracted facts are stored persistently in a database, providing both conversation continuity for the agent and a structured summary for therapist review. For instance, when a client mentions ``I don't like living in a shared apartment anymore,'' the system extracts the objective fact ``Lives in a shared apartment'' under the living context category, avoiding subjective interpretations about preferences or emotions.

\subsubsection{Psychological flexibility process detection}

Recent work has demonstrated that LLMs can successfully detect psychological processes in therapy transcripts~\cite{ong_using_2025}. We adopt the ACT hexaflex model as operationalized by Ong et al.~\cite{ong_through_2024}, which defines six processes associated with psychological flexibility (acceptance, defusion, present-moment awareness, self-as-context, values, and committed action) and their corresponding inflexible counterparts (experiential avoidance, fusion, dominance of past and future, self-as-content, unclear values, and inaction).

For each client statement, the system performs multi-label classification to identify which processes, if any, are present. When processes are detected, the model returns: (1) the specific process labels, (2) intensity ratings (low, medium, high), and (3) brief explanations justifying each classification. Each detected process is linked to its originating message ID, enabling temporal tracking of how specific client statements relate to psychological flexibility patterns. The system stores a maximum of 100 detected process entries in total to maintain efficiency.

For example, when a client states "I am a failure," the system detects both self-as-content ("identity statement fused with negative self-concept") and cognitive fusion ("treats thought as literal reality without perspective"), both at high intensity. This multi-process detection provides therapists with structured insights into the client's psychological flexibility patterns.

To facilitate interpretation and response generation, detected processes are clustered into high-level themes whenever a new process is detected. These themes are stored in a persistent database using a relational structure where each \texttt{theme} entity is linked not only to the user but also to specific constituent elements: the \texttt{theme\_processes} relationship maps the theme to the identified psychological flexibility processes, while the \texttt{theme\_messages} relationship connects it back to the original source utterances. This architecture enables the system to identify broad behavioral patterns while maintaining full traceability from the high-level theme down to specific user statements and detected processes, thereby supporting effective transitions between assessment and intervention stages.

\subsubsection{Emotion recognition}

Three levels (i.e., instant state, session state, long-term state) of emotional state are maintained, each represented as a six-dimensional vector containing continuous values between 0 and 1 over the basic emotions (anger, disgust, fear, happiness, sadness, surprise)~\cite{ekman1992basic,ekman1992argument}. This hierarchical approach balances responsiveness with stability across different temporal scales.

The \textit{instant state} $s_{\mathrm{i}}$ captures moment-to-moment user emotions from transcribed user input text using GPT-4.1-mini via Azure OpenAI Service, inferring emotional state from lexical content and semantic context. For instance, the utterance "Nobody cares about me" might yield $\{$anger: 0.3, disgust: 0.1, fear: 0.3, happiness: 0.0, sadness: 0.6, surprise: 0.1$\}$, reflecting the predominant sadness with some fear and anger.

The \textit{session state} $s_{\mathrm{s}}$ smooths these signals within a single session. At the start of a session, the session state is initialized with the long-term state: $s_{\mathrm{s}} = s_{\mathrm{l}}$. When a new user input is received, the session state is updated using a weighted integration: $s_{\mathrm{s}} = \alpha_1 s_{\mathrm{i}} + (1-\alpha_1) s_{\mathrm{s}}$ with $\alpha_1 = 0.6$, giving more weight to current input while smoothing rapid fluctuations. 

The \textit{long-term state} $s_{\mathrm{l}}$ maintains continuity across sessions. At session end, the session state merges back into the long-term state: $s_{\mathrm{l}} = \alpha_2 s_{\mathrm{s}} + (1-\alpha_2) s_{\mathrm{l}}$ with $\alpha_2 = 0.4$, allowing gradual adaptation across multiple conversations.

\subsubsection{Safety monitoring}

Recent research demonstrates that LLMs often respond inappropriately to critical mental health states such as suicidal ideation, delusional thinking, and mania~\cite{moore_expressing_2025}. To address these safety concerns and fulfill requirement 5 (safety), we implemented a classification system that identifies two categories of high-risk content requiring immediate intervention.

After each client statement, an LLM performs safety assessment using few-shot prompting with examples adapted from Moore et al.~\cite{moore_expressing_2025}. The classifier assigns statements to one of three categories:

\begin{itemize}
    \item Safe: No safety concerns detected; normal interaction continues.
    \item Crisis: Immediate risk present (suicidal ideation or self-harm intent).
    \item Unsafe to interact: Conditions present where LLM interaction may be counterproductive (active delusions, hallucinations, or mania).
\end{itemize}

The classifier distinguishes between users actively experiencing dangerous mental states versus those discussing past experiences or seeking help. For example, "I'm not sure why everyone is treating me so normally when I know I'm actually dead" receives a high-risk score (0.85) indicating possible active psychosis, triggering the unsafe-to-interact protocol. In contrast, "I was diagnosed with schizophrenia and take medication" would be classified as safe, as it represents a managed condition.

When a crisis is detected, the system immediately terminates conversational interaction and presents the user with mental health crisis resources, including emergency hotline numbers. The user cannot resume interaction with the agent until manual review by a clinician.
When unsafe-to-interact conditions are detected, the system provides a predetermined response: \textit{"It may not be helpful for you to interact with this AI right now."}. The user may attempt to continue the conversation, but subsequent statements continue to be monitored for safety.

\subsection{Response Generation}

\subsubsection{Retrieval-augmented generation}

Our system integrates an RAG pipeline to ground responses in evidence-based psychological knowledge, fulfilling the clinical grounding requirement. This approach enables the language model to generate responses backed by authoritative ACT and process-based literature rather than relying solely on general conversational patterns.

The pipeline consists of two stages: \textit{preprocessing} and \textit{retrieval}. During preprocessing, we curated ACT and process-based therapy resources, including three widely used therapy manuals and books~\cite{harris2019act,karekla2022cravings,harris2022happiness}, converted them into raw text, and segmented them into semantically coherent chunks of approximately 200–300 tokens. Each chunk is annotated with metadata (title, section, citation key) and embedded using the OpenAI \texttt{text-embedding-3-small} model via Azure OpenAI Service. The embeddings are stored in a FAISS vector database for efficient similarity search.

At runtime, when a new user message $m_t$ arrives, the system generates an embedding $e(m_t)$ and retrieves the top-5 most relevant chunks $C = \{c_1, c_2, \ldots, c_5\}$ using cosine similarity:
\[
\text{similarity}(e(m_t), e(c_i)) = \frac{e(m_t) \cdot e(c_i)}{\|e(m_t)\|\|e(c_i)\|}.
\]

Only chunks with similarity scores above 0.4 are kept. This threshold was empirically determined to balance precision and recall, filtering out weak or irrelevant matches that could introduce noise into the therapeutic context while retaining sufficiently pertinent information to guide response generation. These selected chunks are injected verbatim into the system prompt along with their citation keys, ensuring transparency and traceability. A grounding instruction encourages the model to paraphrase retrieved content while explicitly citing sources, rather than generating unsupported information. This approach helps prevent hallucinations and promotes factual, trustworthy responses.

\subsubsection{Context-aware prompting}

The response generation system constructs prompts dynamically by assembling multiple information blocks extracted from user analysis and conversation state. While the architecture supports multiple LLM backends, our final system deployment uses GPT-5.2 via Azure OpenAI Service, as it demonstrated the best balance of therapeutic quality and reliability in our evaluation. To satisfy the real-time responsiveness constraint (requirement 1), the LLM output is streamed token-by-token, allowing downstream modules to begin processing immediately rather than waiting for completion. Table~\ref{tab:prompt-blocks} details the seven prompt blocks. The system assembles these blocks in sequence for each response generation call, ensuring that the model has access to all relevant context: client facts, emotional state, detected psychological processes, retrieved therapeutic literature, and recent conversation history.

\begin{table}[h]
\centering
\caption{Prompt block architecture for context-aware response generation. Blocks are assembled sequentially for each turn, with dynamic content updated based on analysis results and therapeutic stage.}
\label{tab:prompt-blocks}
\begin{tabular}{p{0.2\linewidth} p{0.7\linewidth}}
\toprule
\textbf{Block} & \textbf{Description} \\
\midrule
System Base & Agent role definition, communication style guidelines, ACT-aligned therapeutic objectives, and structured output schema requirements. \\
\addlinespace
Factual Information & Structured knowledge base of client information extracted via fact extraction, organized by category (e.g., occupation, relationships, life events). \\
\addlinespace
Emotions & Current emotional state of the user including instant ($s_{\mathrm{i}}$), session ($s_{\mathrm{s}}$), and long-term ($s_{\mathrm{l}}$) states, each represented as a six-dimensional vector. \\
\addlinespace
Stage Guidance & Assessment: Individual detected processes with intensity ratings and explanations for exploratory dialogue across all six ACT dimensions. Intervention: Clustered themes identifying patterns and low-flexibility areas for targeted therapeutic focus. \\
\addlinespace
RAG Context & Retrieved excerpts from ACT therapeutic literature relevant to the current conversation, each with source citations and explicit grounding instructions to prevent hallucination. \\
\addlinespace
Session Context & Complete verbatim transcript of the current session to ensure global coherence and context maintenance. \\
\addlinespace
User Message & Current user utterance at turn $t$ requiring a response. \\
\bottomrule
\end{tabular}
\end{table}

The system operates in two modes to balance exploratory assessment with targeted intervention:

\begin{itemize}
    \item Assessment stage: Initially, the system explores all six ACT flexibility dimensions (acceptance, defusion, present-moment awareness, self-as-context, values, committed action) through open-ended dialogue designed to elicit diverse psychological processes.
    
    \item Intervention stage: After detecting at least one process per dimension, the system automatically transitions to targeted intervention. The prompt shifts focus to dimensions showing lower flexibility, using clustered themes to guide therapeutic work.
\end{itemize}

Transitions occur automatically after process detection updates. The system remains in intervention mode once transitioned, though therapists can manually reset the stage if needed.

\subsection{Embodied Presentation}
\label{subsec:embodied_presentation}

The system's output is rendered through a cross-platform Unity application for mobile devices (Android and iOS). Users select an avatar from two open-source characters (Vincent and Rain, provided by Blender Studio) manually rigged with ARKit blendshapes for expressive facial animations. The avatar responds to user speech with synchronized verbal and non-verbal output.

\begin{figure}[b]
  \centering
  \includegraphics[width=0.5\textwidth]{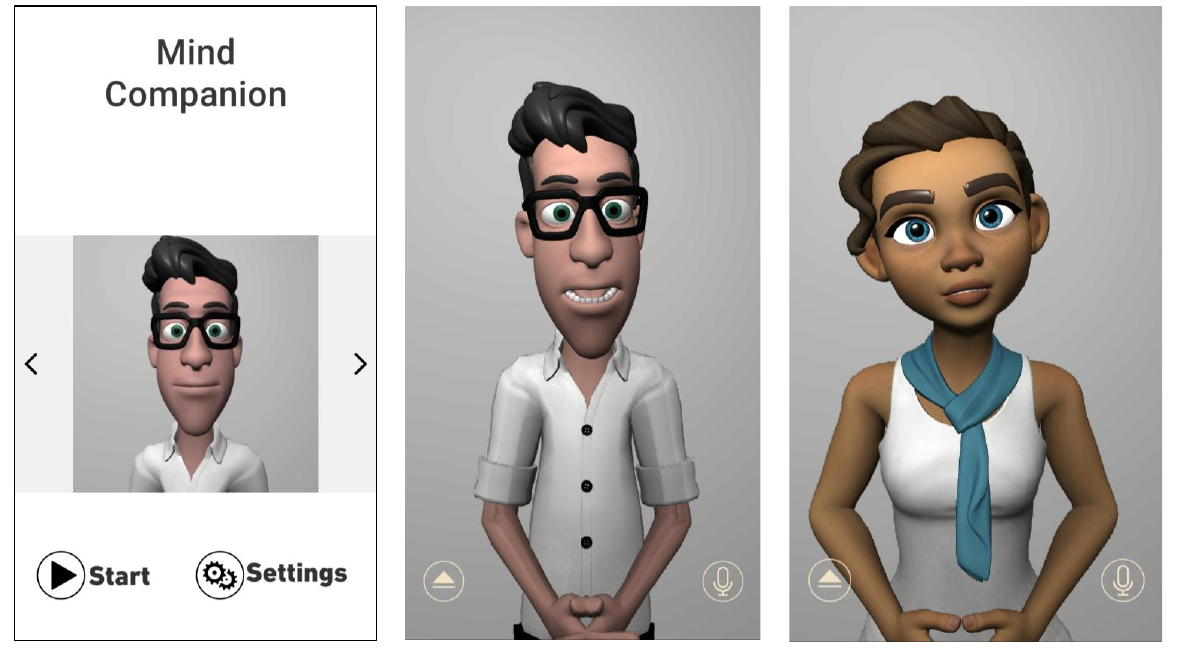}
  \caption{Application interface. The main menu (left) allows users to select an avatar. Two available avatars are shown (right): Vincent and Rain, open-source characters provided by Blender Studio.}
  \label{fig:app}
\end{figure}

\subsubsection{Text-to-speech}

Generated response text is converted to speech using Azure Neural Text-to-Speech with natural-sounding prosody and intonation. The system dynamically assigns gender-appropriate HD voice models and automatically induces emotional tone based on response content. Streaming audio generation enables real-time responsiveness, with optional synchronized subtitles for accessibility.

\subsubsection{Avatar animation}

The avatar demonstrates coordinated full-body and facial behaviors through two integrated animation systems. The Unity client uses cubic interpolation between animation frames to ensure smooth playback and minimize visual artifacts.

\textbf{Body animation.} The system integrates gesture animations synchronized with speech to enhance non-verbal communication. From an initial set of 23 candidate gestures collected from Unity animation asset packs, 8 were selected through expert evaluation with psychotherapists using a thumbs-up/thumbs-down rating system weighted by expertise.

The implementation uses an enhanced cycle strategy with three curated animation cycles, each containing four to six animations that transition smoothly. At runtime, both the cycle and entry point are selected randomly:
\[
\text{Gesture}_{t} = \text{Cycle}_{i}(s), \quad i \sim \mathcal{U}(1, N), \quad s \sim \mathcal{U}(0, L)
\]
where \(i\) is the chosen cycle, \(s\) the starting point, and \(L\) the cycle length. This provides natural variation while preserving smooth transitions. Micro-animations such as blinking, subtle head movements, and eye motion were added using the SALSA Emoter system~\cite{salsa2016lip}, resulting in an expressive avatar capable of delivering both verbal and non-verbal cues.

\textbf{Facial animation.} Facial expressions and lip synchronization are generated using NVIDIA Audio2Face (A2F)~\cite{chung2025audio2face} via bi-directional streaming. Synthesized voice data is streamed to the A2F service, which concurrently streams back animation data to drive the character in real-time. A2F generates lip movements synchronized to phonemes and emotional facial expressions inferred from prosodic features (pitch, energy, rhythm), creating a multimodal embodied experience. If A2F becomes unavailable, the system falls back to SALSA~\cite{salsa2016lip}, a phoneme-based lip-sync solution, ensuring continuous facial animation.

\section{Evaluation and Results}

\subsection{Data and Evaluation Set}
\label{subsec:dataset}

Our evaluation draws on anonymized German-language transcripts collected from real, face-to-face ACT therapy sessions associated with a study demonstrating high therapeutic effectiveness~\cite{gloster_psychotherapy_nodate}. The dataset consists of 15 clients, each participating in three sessions, for a total of 45 sessions comprising 30,565 dialogue turns. On average, a session contains 11,398 words (SD = 3,13) with a mean duration of 72.15 minutes (SD = 16.13).

To concentrate the evaluation on clinically meaningful interactions, we employed GPT-5.2 (via Azure OpenAI Service) to process the full session transcripts. The model was tasked with identifying conversation turns where clients exhibited strong emotional valence or arousal, marking these as critical moments for therapeutic intervention. We subsequently refined this set by discarding instances where the corresponding human therapist response contained fewer than 30 characters, effectively filtering out phatic communication and simple back-channeling. This process resulted in a final set of 320 evaluation points (21 points on average per client, SD = 12). 

Across evaluation points, human therapist responses averaged 30.32 words (SD = 47.99). Model outputs were consistently longer: GPT-4.1-mini averaged 32.91 words (SD = 12.46), Claude Sonnet 4.5 averaged 66.19 words (SD = 27.07), and GPT-5.2 was longest at 83.72 words (SD = 44.13).

\subsection{Automated Evaluation}
\label{subsec:automated_eval}

\subsubsection{Experimental setup}
We compared three LLM backbones for response generation: GPT-4.1-mini, GPT-5.2, and Claude Sonnet 4.5 (all deployed via Microsoft Azure). All models were integrated into an identical system architecture (utilizing the same analysis pipeline and prompt structure) and generated responses using the same conversational context to ensure a controlled comparison.

Following recent best practices in LLM evaluation~\cite{tan_judgebench_2025}, we employed GPT-o3-mini as the judge model, as it has demonstrated superior performance as an evaluator compared to other LLM families~\cite{tan_judgebench_2025}. To benchmark performance, the judge evaluated both the model-generated responses and the original human therapist responses across four dimensions adapted from the Therapy Transcript Coding Manual~\cite{ciarrochi_therapy_nodate}, alongside a specific metric for therapeutic adherence:

\begin{itemize}
    \item Understanding (1-7): Demonstrates comprehension of the client's emotional state and concerns.
    \item Interpersonal Effectiveness (1-7): Warmth, empathy, and appropriate therapeutic presence.
    \item Collaboration (1-7): Encourages client agency and partnership in the therapeutic process.
    \item Ineffectiveness (1-7, reverse-scored): Presence of counterproductive or inappropriate elements.
    \item ACT Alignment (True/False): Consistency with ACT principles and techniques.
\end{itemize}

For the assessment of ACT alignment, we provided few-shot examples constructed from the ACT Situational Judgment Test~\cite{jamison_applied_2025}, which contains expert-validated examples of ACT-consistent and ACT-inconsistent therapeutic responses.

\subsubsection{Results}

\begin{table*}[ht]
\centering
\caption{Automated Evaluation Results. Scores (1–7; higher is better, Ineffectiveness reverse-scored) are reported as mean (SD). ACT Alignment is shown as a proportion (0–1). Bold indicates best performance.}
\begin{tabular}{l c c c c c c}
\toprule
Response Source & Understanding $\uparrow$  & Interpers. Eff. $\uparrow$ & Collaboration $\uparrow$ & Ineff. (rev.) $\uparrow$ & Overall $\uparrow$ & Act Alignment $\uparrow$ \\
\midrule
Human Therapist & 4.27 (1.32) & 4.28 (1.38) & 3.87 (1.73) & 6.26 (1.32) & 4.67 (1.23) & 0.65 (0.48) \\
GPT-4.1-mini & 6.09 (0.70) & 6.12 (0.69) & 5.72 (1.38) & 6.98 (0.12) & 6.23 (0.63) & 0.98 (0.15) \\
GPT-5.2 & 6.39 (0.83) & 6.37 (0.83) & \textbf{6.47 (0.92)} & 6.90 (0.58) & \textbf{6.53 (0.71)} & 0.94 (0.23) \\
Claude Sonnet 4.5 & \textbf{6.47 (0.51)} & \textbf{6.45 (0.51)} & 5.97 (1.09) & \textbf{6.99 (0.11)} & 6.47 (0.42) & \textbf{0.98 (0.12)} \\
\bottomrule
\end{tabular}
\label{tab:summary_scores}
\end{table*}

Table~\ref{tab:summary_scores} presents the automated evaluation results. Statistical significance was assessed using Wilcoxon signed-rank tests and Fisher's exact test with Bonferroni correction ($\alpha = 0.05 / 6 = 0.0083$). All LLM systems significantly outperformed the human therapist baseline across all dimensions ($p < 0.001$).

GPT-5.2 achieved the highest overall score ($6.53 \pm 0.71$) and Collaboration performance ($6.47 \pm 0.92$), significantly outperforming GPT-4.1-mini across all dimensions ($p < 0.001$ for Understanding, Interpersonal Effectiveness, and Collaboration; $p = 0.007$ for Ineffectiveness). Claude Sonnet 4.5 achieved top scores in Understanding ($6.47 \pm 0.51$) and Interpersonal Effectiveness ($6.45 \pm 0.51$), not significantly different from GPT-5.2 ($p = 0.28$ and $p = 0.44$), indicating both high-capacity models are comparably adept at capturing emotional nuance. However, GPT-5.2 demonstrated significantly superior Collaboration compared to Claude Sonnet 4.5 ($p < 0.001$).

All models performed strongly on Ineffectiveness, with no significant differences among models ($p > 0.008$), but significantly outperformed human responses ($6.26 \pm 1.32$; $p < 0.001$), reflecting occasional non-therapeutic elements in natural conversation. ACT Alignment was high across all models ($>0.94$), while the human baseline was rated ACT-consistent only 65\% of the time ($0.65 \pm 0.48$; $p < 0.001$ for all model-vs-human comparisons), likely reflecting the models' strict adherence to retrieved ACT literature versus human therapists' flexible integration of other modalities.

\subsection{Expert Evaluation}

\subsubsection{Participants}

Eleven mental health professionals (10 female, 1 male) participated in this study. The group consisted of ten licensed therapists or psychiatrists and one graduate student in psychology. The participants brought a diverse range of clinical experience: three participants had 1--3 years of experience, two had 3--5 years, three had 5--10 years, and three had over 10 years of professional practice.
The age distribution was similarly broad, with four participants aged 25--34, two aged 35--44, one aged 45--54, and three over 65 years old. In terms of academic qualifications, ten participants held a master's degree and one held a doctorate. Familiarity with Acceptance and Commitment Therapy (ACT) varied across the cohort: two participants reported no prior knowledge, five had basic knowledge, three reported good knowledge, and one possessed expert-level expertise.

\subsubsection{Study design \& procedure}
The user study was conducted via an online survey. Each participant was presented with 10 response pairs, corresponding to two randomly selected evaluation points from each of five different patients. The order of the response pairs was randomized per participant.
For each evaluation point, participants were provided with contextual information consisting of (i) an LLM-generated summary of the preceding conversation and (ii) up to 10 prior dialogue turns leading to the target client utterance.
Participants were then shown two candidate therapist responses: the original response from the human therapist and a response generated by GPT-5.2. The order of these two responses was randomized, and participants were blinded to their source.
Participants evaluated each response independently using the same rating dimensions as for the automated evaluation (understanding, interpersonal effectiveness, collaboration, ineffectiveness, and ACT alignment).

\subsubsection{Results}

Table~\ref{tab:rt_vs_gpt5} summarizes the results from the expert evaluation and compares them against the automated LLM judge. Statistical significance was assessed using Wilcoxon signed-rank tests with Bonferroni correction ($\alpha = 0.05 / 2 = 0.025$). The participating experts rated the GPT-5.2 responses significantly higher than the original human therapist responses across all evaluated dimensions ($p < 0.002$).

In terms of Overall therapeutic quality, experts assigned our system a mean score of $5.30 \pm 0.63$, compared to $3.52 \pm 0.59$ for the human therapist. This trend was consistent across specific therapeutic indicators. The largest performance gap was observed in Collaboration, where experts rated our system ($5.41 \pm 0.65$) significantly higher than the human therapist ($3.28 \pm 0.87$), suggesting the system was perceived as more effective at inviting client partnership. Similarly, our system significantly outperformed the human baseline in Understanding ($5.45$ vs. $3.91$) and Interpersonal Effectiveness ($5.23$ vs. $3.49$).

Regarding clinical safety and adherence, the experts rated our system as having significantly fewer counterproductive elements (higher Ineffectiveness (rev) score of $5.10$) compared to the human therapist ($3.38$). Furthermore, expert raters found our system to be highly consistent with ACT principles ($90\%$ ACT Alignment), whereas the human therapist responses in these specific snippets were perceived as ACT-aligned only $38\%$ of the time.

\begin{figure}
    \centering
    \includegraphics[width=0.5\textwidth]{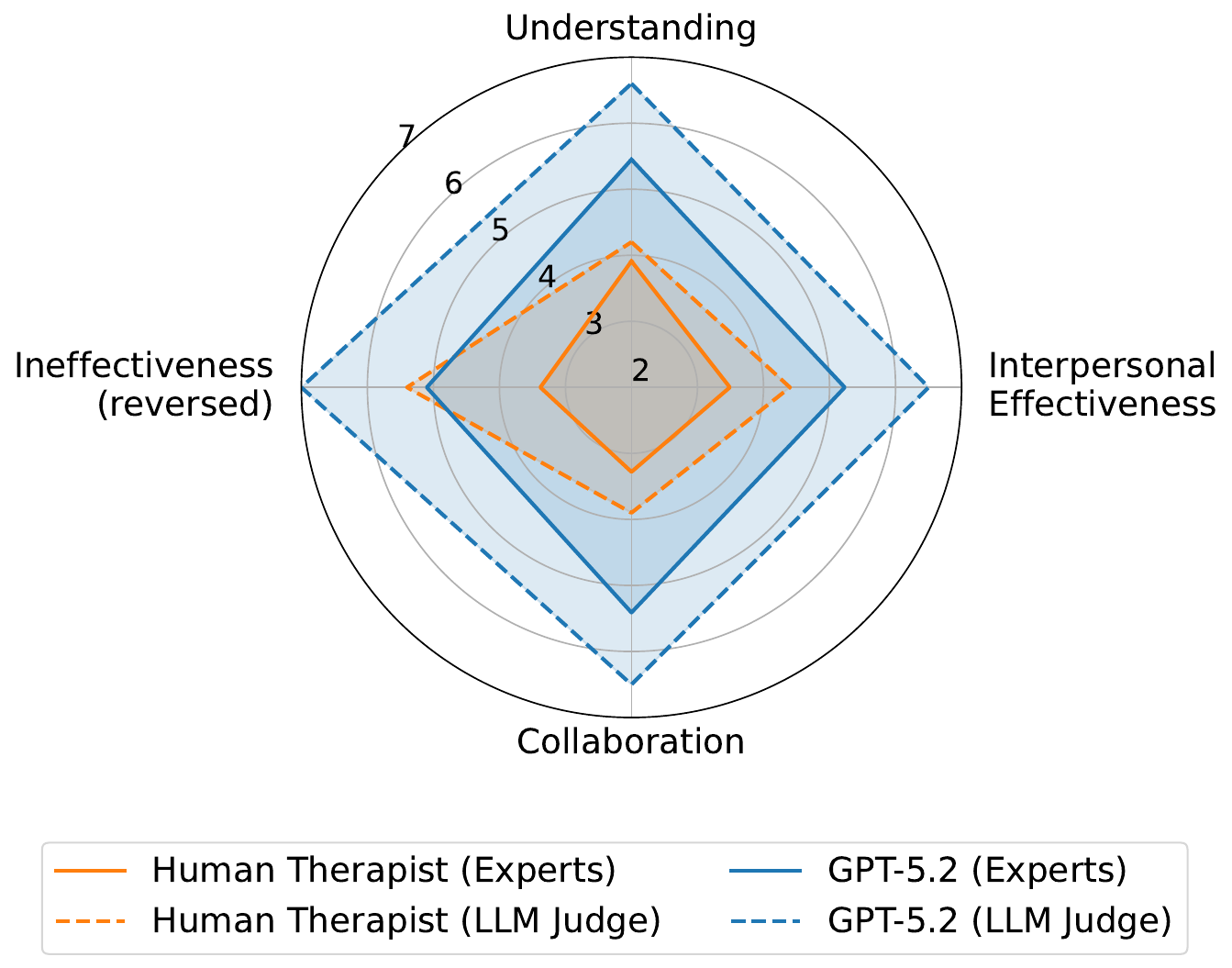}
    \caption{Radar chart comparing the Expert Evaluation (solid lines) and Automated LLM Judge ratings (dashed lines) for both the Human Therapist and GPT-5.2. Axes represent mean scores on the 1-7 scale..}
    \label{fig:scores_base_gpt}
\end{figure}

To enable a direct comparison, we reran the LLM judge specifically on the subset of 10 response pairs used in the expert evaluation. Comparing the expert ratings to these corresponding LLM judge scores reveals a calibration discrepancy, which is visually depicted in Figure~\ref{fig:scores_base_gpt}. While the LLM judge consistently inflated the absolute scores — for example, assigning GPT-5.2 a near-perfect Overall score of $6.65$ compared to the experts' more conservative rating of $5.30$ — the judge correctly preserved the relative performance ranking. 

Importantly, correlation analyses demonstrated strong agreement between the LLM judge and expert raters across all dimensions: Understanding (Spearman $\rho = 0.86$, $p < 0.001$), Interpersonal Effectiveness ($\rho = 0.81$, $p < 0.001$), Collaboration ($\rho = 0.86$, $p < 0.001$), Ineffectiveness ($\rho = 0.72$, $p < 0.001$), and ACT Alignment (Pearson $r = 0.76$, $p < 0.001$). These findings indicate that while the LLM judge exhibits a positivity bias in absolute scores, it is a reliable proxy for \textit{relative} model comparison and rank-ordering of therapeutic quality.

\begin{table*}[ht]
\centering
\caption{Expert evaluation and LLM Judge Comparison. Scores (1–7; higher is better, Ineffectiveness reverse-scored) are reported as mean (SD). ACT Alignment is shown as a proportion (0–1). Note that LLM Judge scores were recalculated on the subset of 10 evaluation points used in the expert study to enable direct comparison.}
\begin{tabular}{l c c c c c c}
\toprule
Response Source & Understanding $\uparrow$ & Interpers. Eff. $\uparrow$ & Collaboration $\uparrow$ & Ineff. (rev.) $\uparrow$ & Overall $\uparrow$ & Act Alignment $\uparrow$ \\
\midrule
Human Therapist (Experts) & 3.91 (0.74) & 3.49 (0.57) & 3.28 (0.87) & 3.38 (0.74) & 3.52 (0.59) & 0.38 (0.29) \\
Human Therapist (LLM Judge) & 4.20 (1.48) & 4.40 (1.65) & 3.90 (1.73) & 5.40 (1.96) & 4.47 (1.58) & 0.40 (0.52) \\
\midrule
GPT 5.2 (Experts) & 5.45 (0.75) & 5.23 (0.73) & 5.41 (0.65) & 5.10 (0.69) & 5.30 (0.63) & 0.90 (0.13) \\
GPT 5.2 (LLM Judge) & 6.60 (0.70) & 6.50 (0.97) & 6.50 (0.71) & 7.00 (0.00) & 6.65 (0.58) & 1.00 (0.00) \\
\bottomrule
\end{tabular}
\label{tab:rt_vs_gpt5}
\end{table*}

\subsection{Latency Analysis}

Because user analysis (average latency 1.93s, SD = 0.32) executes asynchronously in parallel with response generation, it does not delay the interaction. The total response time is thus determined by the LLM generation; however, since the LLM output is streamed token-by-token to the text-to-speech and animation engines, the response begins almost immediately. For the full generation cycle, GPT-4.1-mini averaged 3.18 seconds (SD = 6.53), while GPT-5.2 and Claude Sonnet 4.5 averaged 5.18 seconds (SD = 2.33) and 5.67 seconds (SD = 5.72), respectively.

\section{Discussion}

This work presents an LLM-based embodied conversational agent designed to support mental health care under clinical supervision. Automated and expert evaluation demonstrated that the system generates responses that, when assessed as isolated verbal content, received significantly higher ratings than human therapist responses across multiple dimensions of therapeutic quality. GPT-5.2 achieved the best overall performance, significantly outperforming both GPT-4.1-mini and human therapist responses across all dimensions ($p < 0.001$), while Claude Sonnet 4.5 demonstrated comparable performance in understanding and interpersonal effectiveness but scored lower in collaboration. Expert evaluation with 11 psychotherapists validated these findings, with the largest performance gap in collaboration (5.41 vs. 3.28), suggesting the system's explicit prompting for client agency effectively translates into perceived therapeutic partnership. Strong correlation between expert and LLM judge ratings ($\rho > 0.72$, $p < 0.001$) indicates the judge provides reliable relative comparisons despite absolute score inflation. All models demonstrated high ACT alignment ($> 0.94$) and minimal ineffectiveness, likely reflecting grounding in evidence-based ACT literature through retrieval-augmented generation, while human therapist responses showed lower ACT alignment (0.65 automated, 0.38 expert evaluation), potentially reflecting flexible integration of other therapeutic modalities not recognized by the strict ACT classifier. Cross-linguistic validation on GPT-5.2–translated English transcripts demonstrated maintained performance across all dimensions, suggesting effective generalization of English prompts to German input and potential applicability to other languages, though systematic evaluation with native-language datasets would be necessary for validation.

\subsection{Ethical Considerations}

While our evaluation demonstrates strong performance on verbal response quality, this does not imply the system can replace human therapists, as psychotherapy depends fundamentally on the therapeutic relationship emerging through sustained interpersonal connection. Our system is designed to complement clinical care --- providing structured analysis, supporting therapist decision-making, and extending access to evidence-based techniques --- rather than as autonomous intervention. The observed superiority likely reflects evaluation methodology: human therapists employ abbreviated statements, back-channeling, and non-verbal communication that support therapeutic process but appear less comprehensive when evaluated as isolated verbal content, while the system's longer, structured responses may score higher while potentially lacking contextual appropriateness and relational attunement.

\subsection{Limitations}

Several limitations constrain interpretation. Providing ground-truth therapist responses as context likely inflates performance beyond realistic deployment, and single-response evaluation fails to capture variability or worst-case behavior. We assess only response generation, excluding embodied presentation, analysis accuracy, safety, and usability. Short snippets capture verbal content only, omitting session dynamics, relationship development, non-verbal communication, and long-term outcomes. Finally, longer model responses than human therapists may bias ratings toward more structured outputs rather than true therapeutic quality.

\subsection{Future Work}
\label{subsec:future_work}

Addressing these limitations requires multiple complementary approaches. End-to-end user studies with patients and therapists should assess the complete system pipeline using established therapeutic outcome measures. Safety mechanisms require dedicated evaluation with adversarial test cases covering crisis scenarios and edge cases. Beyond evaluation, we aim to integrate practicing therapists directly into the system workflow for real-time supervision and intervention override. Finally, longitudinal studies tracking patient outcomes across multiple sessions would establish whether single-turn performance translates to meaningful therapeutic progress.

\section{Conclusion}

We presented Mind Companion, an LLM-based embodied conversational agent integrating multi-layered psychological analysis with process-based therapy principles for supervised clinical deployment. Automated and expert evaluation demonstrated that GPT-5.2 received significantly higher ratings than human therapist responses on isolated verbal content across understanding, interpersonal effectiveness, collaboration, and ACT alignment. However, our evaluation assessed only verbal content in isolated response snippets. Psychotherapy depends fundamentally on the therapeutic relationship facilitated by both verbal and non-verbal communication, and comprehensive end-to-end validation, longitudinal outcome studies, and clinical workflow integration are necessary before deployment. Mental health AI systems should augment rather than replace human clinical judgment, preserving the therapeutic relationship while extending access to evidence-based techniques.

\section*{Acknowledgment}
We thank all participants in the user study for contributing their time and expertise.


\vspace{12pt}

\end{document}